\documentclass{PoS}

\usepackage{amsmath,bm}
\usepackage{amsmath}

\title{Intrinsic quantum mechanics\\ behind the Standard Model?}

\ShortTitle{Intrinsic quantum mechanics}

\author{\speaker{Ole L. Trinhammer}\\
        Department of Physics, Technical University of Denmark,\\ Fysikvej bld 307, DK2800 Kongens Lyngby, Denmark\\
        E-mail: \email{ole.trinhammer@fysik.dtu.dk}}

\abstract{We suggest the gauge groups SU(3), SU(2) and U(1) to share a common origin in U(3).
We take the Lie group U(3) to serve as an intrinsic configuration space for baryons. A spontaneous symmetry break in the baryonic state selects a U(2) subgroup for the Higgs mechanism. The Higgs field enters the symmetry break to relate the strong and electroweak energy scales by exchange of one quantum of action between the two sectors. This shapes the Higgs potential to fourth order.
Recently intrinsic quantum mechanics has given a suggestion for the Cabibbo angle from theory (EPL124-2018) and a prediction for the Higgs couplings to gauge bosons (EPL125-2019). Previously it has given the nucleon mass and the parton distribution functions for u and d quarks in the proton (EPL102-2013). It has given a quite accurate equation for the Higgs mass in closed form (IJMPA30-2015) and an N and Delta spectrum essentially without missing resonances (arXiv:1109.4732).
The intrinsic space is to be distinguished from an interior space. The intrinsic space is non-spatial, i.e. no gravity in intrinsic space. The configuration variable is like a generalized spin variable excited from laboratory space by kinematic generators: momentum, spin and Laplace-Runge-Lenz operators.
The baryon dynamics resides in a Hamiltonian on U(3) and projects to laboratory space by the momentum form of the wavefunction. The momentum form generates conjugate quark and gluon fields. Local gauge invariance in laboratory space follows from unitarity of the configuration variable and left invariance of the coordinate fields on the intrinsic space.
Future work should aim to invoke leptons in the second and third generations and quarks in the third.
}

\FullConference{
European Physical Society Conference on High Energy Physics - EPS-HEP2019 -\\
			10-17 July, 2019\\
			Ghent, Belgium}

\begin{document}

\section{Intrinsic configuration space for baryons}

In 1925 George Uhlenbeck and Samuel Goudsmit were the first to see the electron spin as a new, intrinsic degree of freedom to explain splitting of atomic levels in magnetic fields \cite{UhlenbeckGoudsmit}.

With a generalization of the idea of intrinsic variables, we consider baryons as stationary states on a compact $U(3)$ Lie group configuration space \cite{TrinhammerEPL102}
\begin{equation}	\label{eq:schroedingerU3}
 \frac{\hbar c}{a}\left[-\frac{1}{2}\Delta+\frac{1}{2}{\rm Tr}\ \chi^2\right]\Psi(u)={\cal E}\Psi(u),\ \ \ u=e^{i\chi}\in U(3).
\end{equation}
The configuration variable $u$ is generated by nine kinematic generators $T_j,S_j,M_j$, thus
\begin{equation}	\label{eq:chiExpandedAndIntrinsicMomentumDefined}
 \chi=\left(a\theta_jp_j+\alpha_jS_j+\beta_jM_j)\right)/\hbar,\ \ \ \ \ \ \ p_j\equiv -i\hbar\frac{1}{a}\frac{\partial}{\partial\theta_j}=\frac{\hbar}{a}T_j,\ \ \ \theta_j,\alpha_j,\beta_j\in\mathbb{R}, j=1,2,3
\end{equation}
where $e^{i\theta_j}$ are the three eigenvalues of $u$ and $\theta_j$ are dynamical angular variables. We interpret the three toroidal generators $iT_j$ as colour generators. The toroidal generators are conjugate to the angular variables, i.e.
\begin{equation}	\label{eq:actionAngleQuantization}
 [iT_j,\theta_i]=\delta_{ij}\sim{\rm d}\theta_i(\partial_j)=\delta_{ij},\ \ \      \ \ \ \partial_j|_u=uiT_j,
\end{equation}
where $\partial_j$ are left invariant coordinate fields on $U(3)$ and ${\rm d}\theta_j$ are corresponding coordinate forms, see e.g. pp. 84 in \cite{Warner}.

\begin{figure}
\includegraphics[width=.35\textwidth]{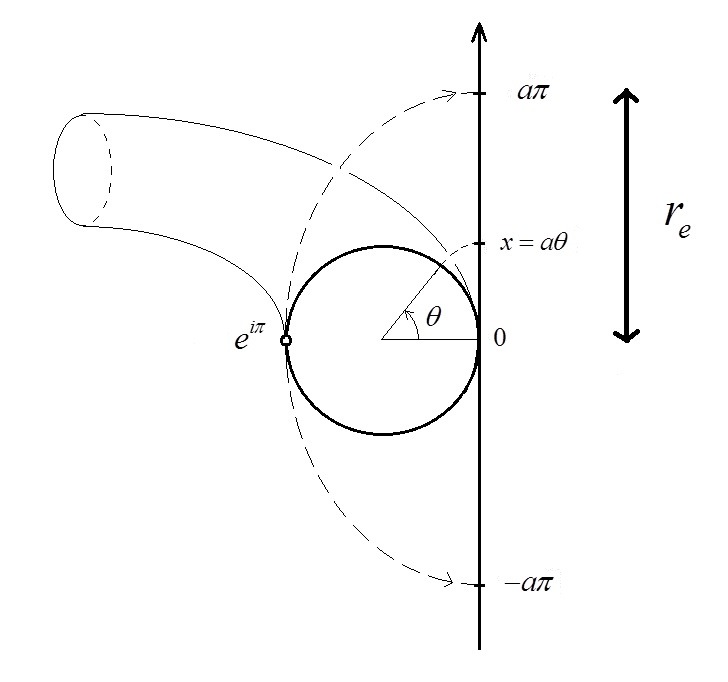}
\includegraphics[width=.64\textwidth]{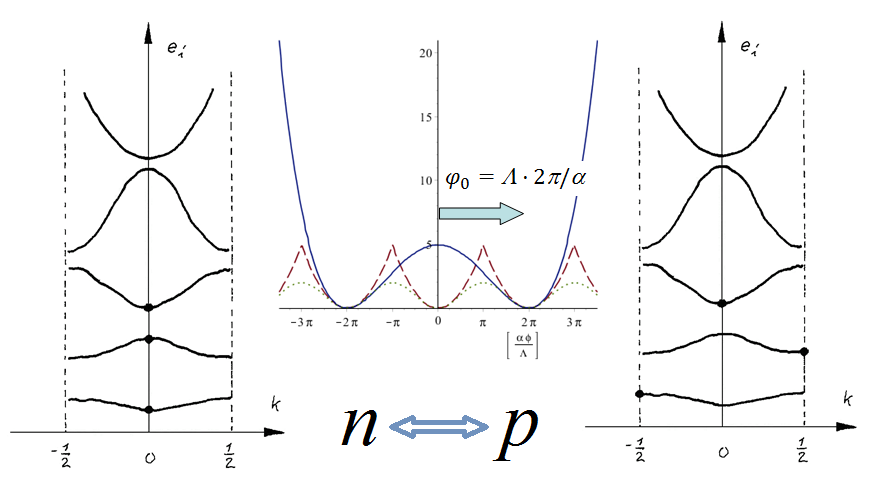}
\caption{{\it First left}: The dynamical toroidal angles of the intrinsic configuration space $U(3)$ maps to laboratory space coordinates via a length scale $a$ determined as $\pi a=r_e$ from the classical electron radius $r_e$. The classical electron radius is determined from the classical electrostatic "self-energy" of the electron, $\frac{e^2}{4\pi\varepsilon_0r_e}=m_ec^2$, p. 97 in \cite{LandauLifshitz}. The intrinsic potential folds out to be periodic in the angular variables, see third. {\it Third from left}: The Higgs potential ({\it solid, blue}) is shaped and scaled by the intrinsic potential ({\it dashed, red}). The potential is inherited from Manton's action in lattice gauge theory \cite{Manton}. We also show a potential inspired by Wilson's action ({\it dotted, green}) \cite{Wilson}. The two potentials give the same fit and thus the same value for Higgs mass and self-coupling. They also give the same constant term and thus the same value for dark energy to baryon matter ratio \cite{TrinhammerIQM4researchGate}, but only the Manton-inspired potential gives a satisfactory reproduction of the baryon spectrum \cite{TrinhammerArxiv1109-4732v3}. The reduced zone schemes, p. 160 in \cite{AshcroftMermin}, ({\it second and fourth from left}) show choices of Bloch wave vectors $\kappa$ for neutronic and protonic states respectively. Figure left from \cite{TrinhammerBohrJensen} and right from \cite{TrinhammerIQM4researchGate}.}
\label{fig1}
\end{figure}

The scale $a$ enters the model when the three toroidal degrees of freedom are related to the three spatial degrees of freedom in laboratory space by the projection, see fig. \ref{fig1}
\begin{equation}	\label{eq:spaceProjection}
 x_j=a\theta_j.
\end{equation}
Thus we define intrinsic momentum operators $p_j$ proportional to the toroidal generators as stated in (\ref{eq:chiExpandedAndIntrinsicMomentumDefined}). Hereby the canonical action-angle quantization in (\ref{eq:actionAngleQuantization}) translates into momentum-position quantization at the {\it origo} of the configuration space
\begin{equation}	\label{eq:positionMomentumQuantization}
 [p_j,a\theta_i]=-i\hbar\delta_{ij}.
\end{equation}
It is via (\ref{eq:spaceProjection}) and (\ref{eq:positionMomentumQuantization}) that the physical dimensions enter the model (\ref{eq:schroedingerU3}). 

The $S_j$ we interpret as intrinsic spin and $M_j$ contains flavour. The $M_j$s commute like Laplace-Runge-Lenz operators and the $S_j$s commute as intrinsic spin operators like in body fixed coordinates in nuclear physics (see e.g. p. 87 in \cite{BohrMottelson}) 
\begin{equation}
 \left[M_i,M_j\right]=\left[S_i,S_j\right]=-i\hbar\varepsilon_{ijk}S_k.
\end{equation}
In a coordinate representation (see e.g. pp. 210 in \cite{Schiff}) we have e.g. \cite{TrinhammerEPL102}
\begin{equation}
 S_1=a\theta_2p_3-a\theta_3p_2=\hbar\lambda_7,\ \ \ \ M_1/\hbar=\theta_2\theta_3+\frac{a^2}{\hbar^2}p_2p_3=\lambda_6.
\end{equation}
Here the Gell-Mann matrices $\lambda_7$ and $\lambda_6$ represent the generators in a $3\times3$ matrix representation (see e.g. pp. 209 in \cite{Schiff}). The trace potential, (see fig. \ref{fig1})
\begin{equation}	\label{eq:intrinsicPotential}
 \frac{1}{2}{\rm Tr}\ \chi^2=\sum_{j=1}^3w(\theta_j),
\end{equation}
where
\begin{equation}
 w(\theta)=\frac{1}{2}\left(\theta-n\cdot2\pi\right)^2,\ \ \ \theta\in\left[(2n-1)\pi,(2n+1)\pi\right],\ \ \ n\in\mathbb{Z}
\end{equation}
depends only on the three eigenvalues $e^{i\theta_j}$ of $u$ since the trace is invariant under conjugations $u\rightarrow v^{-1}uv,v\in U(3)$ and in particular under diagonalization of $u$. The potential is periodic in the three {\it eigenangles} $\theta_j, j=1,2,3$ manifesting the compactness of the configuration space. The Laplacian in a polar decompostion reads \cite{TrinhammerOlafsson}
\begin{equation}	\label{eq:laplacianAndJacobian}
 \Delta=\sum_{j=1}^3\frac{1}{J^2}\frac{\partial}{\partial\theta
 _j}J^2\frac{\partial}{\partial\theta
 _j}-\sum_{\substack{i<j\\ k\neq i,j}}^3\frac{\left(S_k^2+M_k^2\right)/\hbar^2}{8\sin^2\frac{1}{2}(\theta_i-\theta_j)},\ \ \ \ \ \ J=\prod_{i<j}^32\sin\left(\frac{1}{2}\left(\theta_i-\theta_j\right)\right).
\end{equation}
The measure-scaled wavefunction $\Phi=J\Psi$ can be factorized into
\begin{equation}	\label{eq:wavefunctionFactorized}
 \Phi(u)=R(\theta_1,\theta_2,\theta_3)\Upsilon(\alpha_1,\alpha_2,\alpha_3,\beta_1,\beta_2,\beta_3).
\end{equation}

The quantum fields on which QCD \cite{RPP2018} is based, are generated by the momentum form of the measure-scaled wavefunction acting on quark and gluon generators from the Lie group algebra $T_j,\ j=1,2,3$ and $\lambda_k,\ k=1,2,\cdots,8$ respectively, see sec. \ref{sec:localGaugeInvariance}.

The off-diagonal variables $\alpha_j, \beta_j$ in  (\ref{eq:chiExpandedAndIntrinsicMomentumDefined}), (\ref{eq:wavefunctionFactorized}) can be integrated out and the Schr\"odinger equation (\ref{eq:schroedingerU3}) solved by a Rayleigh-Ritz method \cite{BruunNielsen, TrinhammerIQMbookResearchGate}, expanding the measure-scaled toroidal wavefunction $R$ respectively for neutronic and protonic states on Slater determinants
\begin{equation}	\label{eq:fpqr}
 f_{pqr}=\left|\begin{matrix}
 \cos p\theta_1 & \cos p\theta_2 & \cos p\theta_3\\
 \sin q\theta_1 & \sin q\theta_2 & \sin q\theta_3\\
 \cos r\theta_1 & \cos r\theta_2 & \cos r\theta_3
 \end{matrix}\right|\in R_n\ \ \ {\rm and}\ \ \
 \frac{1}{2}\left(e^{\frac{i}{2}(\theta_1+\theta_2+\theta_3)}+e^{\frac{-i}{2}(\theta_1+\theta_2+\theta_3)}\right)f_{pqr}\in R_p
\end{equation}
where $p,q,r$ are integers choosen to constitute a complete base. The Bloch phase factors in the protonic state signify the creation of electric charge. Initially the eigenangles $\theta_i,\theta_j$ acquire Bloch phases pairwise to ensure integrability of the second (centrifugal) term in the Laplacian in (\ref{eq:laplacianAndJacobian}). We take this pairing to frame the $U(2)$ structure of the Higgs mechanism and we use the Ansatz \cite{TrinhammerArxivHiggsMass}
\begin{equation}	\label{eq:trailingAnsatz}
 \Lambda 2\pi=\alpha\varphi_0\ \ \ \sim\ \ \ \hbar c2\pi=\alpha\varphi_0a
\end{equation}
to relate the strong and electroweak sectors by {\it exchange of one unit of space action} $hc$ and together with the intrinsic potential (\ref{eq:intrinsicPotential}) to scale and shape the Higgs potential \cite{TrinhammerBohrJensen}
\begin{equation}	\label{eq:higgsPotential}
 V_H(\phi)=\frac{1}{2}\delta^2\phi_0^2-\frac{1}{2}\mu^2\phi^2+\frac{1}{4}\lambda^2\phi^2,\ \ \ \ \ \ \ \delta^2=\frac{1}{4}\varphi_0^2,\ \ \mu^2=\frac{1}{2}\varphi_0^2,\ \ \lambda^2=\frac{1}{2}.
\end{equation}
Here $\Lambda=\hbar c/a$ from (\ref{eq:schroedingerU3}), $\alpha$ is the fine structure coupling and $\varphi_0$ is the Higgs field vacuum expectation value determined by (\ref{eq:trailingAnsatz}).

\section{Local gauge invariance from intrinsic structure}
\label{sec:localGaugeInvariance}

We show here how local gauge invariance follows from left invariant coordinate fields and how unitary configuration variables are required.

First we generate colour quark fields $\psi_j$ by use of the momentum form on the measure-scaled toroidal wavefunction $R$ in (\ref{eq:wavefunctionFactorized}), thus
\begin{equation}	\label{eq:colourFields}
 \psi_j(u)={\rm d}R_u(iT_j)=\partial_j|_u[R].
\end{equation}
In general the momentum form acts on an element $Z$ of the algebra as
\begin{equation}	\label{eq:fieldsGeneral}
 {\rm d}\Phi_u(Z)\equiv Z_u[\Phi]=\frac{\rm d}{{\rm d}\ t}\Phi(ue^{tZ})|_{t=0}
\end{equation}
and is also called a derivation or an exterior derivative. Think of a directional derivative on a curved surface along a tangent direction. When restricting to the torus as in (\ref{eq:colourFields}), we generate colour quark fields. The six off-toroidal derivatives generate six of the eight gluon fields. The remaining two gluon fields correspond to the diagonal Gell-Mann generators $\lambda_3$ and $\lambda_8$ which are linear combinations of the three toroidal generators $T_j$.

Secondly we require local gauge invariance of a Hamiltonian (p. 145 in \cite{Sakurai}) constructed from the colour quark fields generated in (\ref{eq:colourFields})
\begin{equation}	\label{eq:hamiltonianSakurai}
 H=\int\psi^\dagger\left(-i\hbar c\bm{\alpha}\cdot \bm{\nabla}+\beta mc^2\right)\psi\ {\rm d} x^3,\ \ \ \psi^\dagger=(\psi_1^*,\psi_2^*,\psi_3^*).
\end{equation}
Here we suppressed spinor indices which are mixed by the $4\times 4$ Dirac matrices $\bm{\alpha}=(\alpha_1,\alpha_2,\alpha_3$) and $\beta$. The spinor indices commute with the colour indices. Using left invariance of the coordinate fields $\partial_j|_u=uiT_j$ from (\ref{eq:actionAngleQuantization}) we get in the mass term of (\ref{eq:hamiltonianSakurai})
\begin{equation}
 \psi'(u')^\dagger\psi'(u')=\left(u'iT_j[R]\right)^\dagger\left(u'iT_j[R]\right)=\left(iT_j[R]\right)^\dagger(u')^\dagger u'\left(iT_j[R]\right)=\psi(u)^\dagger\psi(u)
\end{equation}
provided the configuration variables $u',u$ are unitary, i.e. $(u')^\dagger u'=u^\dagger u={\bf 1}$. Next we impose the local gauge transformation
\begin{equation}	\label{eq:gaugeTransformation}
 \psi\rightarrow\psi'=g(x)\psi,\ \ g(x)\in SU(3), \ \ \partial_\mu\rightarrow D_\mu=\partial_\mu+A_\mu
\end{equation}
with gauge field expanded on the Gell-Mann matrices $\lambda_k$ including a common strong coupling $g_s$
\begin{equation}
 A_\mu=ig_sA_\mu^k\lambda_k,\ \ k=1,2,\cdots,8
\end{equation}
and the gauge field transforming as (p. 223 in \cite{Scheck})
\begin{equation}	\label{eq:gaugeFieldAndCovariantDerivativeTransformation}
 A_\mu'=g(x)A_\mu g(x)^{-1}+g(x)\partial_\mu g(x)^{-1}\ \ \rightarrow\ \ \left(D_\mu'\psi'\right)^2=\left(D_\mu\psi\right)^2.
\end{equation}
We see that choosing $u=g(x)$ in (\ref{eq:gaugeTransformation}) and in (\ref{eq:colourFields}) equivalences local gauge transformation in laboratory space (\ref{eq:gaugeFieldAndCovariantDerivativeTransformation}) to left translation of the coordinate fields (\ref{eq:actionAngleQuantization}) on the intrinsic configuration manifold.

\section{Some results to compare with observations}

The second and third terms in the Higgs potential (\ref{eq:higgsPotential}) are interpreted as mass and self-coupling terms as usual. We thus get a compact expression for the Higgs mass
\begin{equation}	\label{eq:higgsMass}
 m_Hc^2=\mu=\frac{1}{\sqrt{2}}\varphi_0=\frac{1}{\sqrt{2}}\frac{2\pi}{\alpha(m_W)}\Lambda=\frac{1}{\sqrt{2}}\frac{2\pi}{\alpha(m_W)}\frac{\pi}{\alpha(m_e)}m_ec^2=125.095\pm0.014\ \rm GeV
\end{equation}
to compare with the experimental value $125.10\pm 0.14\ \rm GeV$ \cite{Erler}. In (\ref{eq:higgsMass}) we use $\alpha^{-1}(m_e)=137.035999139(31)$ \cite{RPP2018} and $\alpha^{-1}(m_W)=127.989(14)$ has been obtained by sliding from $\alpha^{-1}(m_Z)=127.950(10)$ \cite{TrinhammerEPL125}.
We also get a prediction for the Higgs self-coupling from (\ref{eq:higgsPotential})
\begin{equation}
 \lambda^2=\frac{1}{2}
\end{equation}
and predict excess Higgs to gauge boson couplings with signal strengths \cite{TrinhammerEPL125}
\begin{equation}	\label{eq:higgsToBosonSignalStrength}
 \frac{\mu_{HVV}}{\mu_{HVV,SM}}=\frac{1}{\left|V_{ud}\right|}\approx 1.03.
\end{equation}
The quark mixing matrix element $V_{ud}$ between $u$ and $d$ quarks enters the expression (\ref{eq:higgsToBosonSignalStrength}) via our determination of the electroweak energy scale (note $G_{F\beta}=G_{F\mu}\left|V_{ud}\right|$ )
\begin{equation}
  v/\sqrt{2}\equiv\varphi_0\ \ \ \rightarrow\ \ v_{\rm SM}=v\sqrt{|V_{ud}|}=246.93(3)\ \rm GeV.
\end{equation}
This is because we set the electroweak scale $v$ from  the neutron decay as seen from (\ref{eq:schroedingerU3}) and (\ref{eq:fpqr}). Note also results on Cabibbo angle \cite{Cabibbo} from $u$ and $s$ quark generators \cite{TrinhammerEPL124} and Weinberg angle from $u$ and $d$ quark generators \cite{TrinhammerEPL124}. And results on proton energy-momentum tensor \cite{TrinhammerEPL128}.

I thank Henrik Georg Bohr and Mogens Stibius Jensen for co-work on the Higgs mass. I thank Jakob Bohr and Steen Markvorsen for helpful discussions on mapping of intrinsic structure to laboratory space.


\begin{thebibliography}{99}

\bibitem{UhlenbeckGoudsmit}
 G. E. Uhlenbeck and S. A. Goudsmit, {\it Ersetzung der Hypothese vom unmechanischen Zwang durch eine Forderung bez\"uglich des inneren Verhaltens jedes einzelnen Elektrons}, Naturw. {\bf 13} (1925) 953-954.
\bibitem{TrinhammerEPL102}
 O. L. Trinhammer, {\it On the electron to proton mass ratio and the proton structure}, EPL {\bf 102} (2013) 42002.
\bibitem{Warner}
 F. W. Warner, {\it Foundations of Differentiable Manifolds and Lie Groups}, Springer, New York 1983.\bibitem{LandauLifshitz}
 L. D. Landau and E. M. Lifshitz, {\it The Classical Theory of Fields. Course of Theoretical Physics}, Volume 2, $4^{\rm th}$ edition, Elsevier-Butterworth-Heinemann, Oxford 2005.
\bibitem{Manton}
 N. S. Manton, {\it An Alternative Action for Lattice Gauge Theories}, Phys. Lett. B {\bf 96} (1980) 328-330.
\bibitem{Wilson}
 K. G. Wilson, {\it Confinement of quarks}, Phys. Rev. D {\bf 10} (1974) 2445-2459.
\bibitem{TrinhammerArxiv1109-4732v3}
 O. L. Trinhammer, {\it Neutron to proton mass difference, parton distribution functions and baryon resonances from dynamics on the Lie group u(3)}, arXiv:1109.4732v3 [hep-th] 25 Jun 2012.
\bibitem{AshcroftMermin}
 N. W. Ashcroft and N. D. Mermin, {\it Solid State Physics}, Holt, Rinehart and Winston, New York 1976.
\bibitem{TrinhammerBohrJensen}
 O. L. Trinhammer, H. G. Bohr and M. S. Jensen, {\it The Higgs mass derived from the U(3) Lie group}, J. Mod. Phys. A {\bf 30} (2015) 1550078. (ArXiv:1503.00620v2 [physics.gen-ph], 7 Dec 2014).
\bibitem{TrinhammerIQM4researchGate}
 O. L. Trinhammer, {\it Intrinsic quantum mechanics IV. Dark energy from Higgs potential}, ResearchGate 25 Jan 2017, DOI:10.13140/RG.2.2.35814.83527, and  O. L. Trinhammer, {\it Dark energy from Higgs potential}, EPL {\bf 130} (2020) 29002.
\bibitem{BohrMottelson}
 Aa. Bohr and B. R. Mottelson, {\it Nuclear Structure. Volume 1. Single Particle Motion}, W. A. Benjamin, New York 1969.
\bibitem{Schiff}
 L. I. Schiff, {\it Quantum Mechanics}, $3^{\rm rd}$ edition, McGraw-Hill, Tokyo, 1968.
\bibitem{TrinhammerOlafsson}
 O. L. Trinhammer and G. Olafsson, {\it The Full Laplace-Beltrami operator on U(N) and SU(N)}, arXiv:math-ph/9901002v2, 12 Apr 2012.
\bibitem{RPP2018}
 M. Tanabashi et al. (Particle Data Group), {\it Review of Particle Physics}, Phys. Rev. D {\bf 38} (2018) 030001.
\bibitem{BruunNielsen}
 Hans Bruun Nielsen, Technical University of Denmark, private communication 1997.
\bibitem{TrinhammerIQMbookResearchGate}
 O. L. Trinhammer, {\it Intrinsic Quantum Mechanics Behind the Standard Model. From Lie group configurations to Particle Physics Observations}, (book) ResearchGate 12 Dec 2018.
\bibitem{TrinhammerArxivHiggsMass}
 O. L. Trinhammer, {\it A Higgs mass at 125 GeV calculated from neutron to proton decay in a u(3) Lie group Hamiltonian framework}, arXiv:1302.1779v2 [hep-ph] 1 May 2014.
\bibitem{Sakurai}
 J. J. Sakurai, {\it Advanced Quantum Mechanics}, Addison-Wesley, Redwood City 1967.
\bibitem{Scheck}
 F. Scheck, {\it Electroweak and Strong Interactions. Phenomenology, Concepts, Models}, $3^{\rm rd}$ edition, Springer, Berlin 2012.
\bibitem{Erler}
 J. Erler, {\it Electroweak Precision Tests of the Standard Model after the Discovery of the Higgs Boson}, Prog. Part. Nucl. Phys. {\bf 106} (2019) 68-199. (ArXiv:1902.05142v2 [hep-ph] 22 Feb 2019).
\bibitem{TrinhammerEPL125}
 O. L. Trinhammer, {\it Excess Higgs to gauge boson couplings}, EPL {\bf 125} (2019) 41001.
\bibitem{Cabibbo}
 N. Cabibbo, {\it Unitary symmetry and leptonic decays}, Phys. Rev. Lett. {\bf 10} (1963) 531-533.
\bibitem{TrinhammerEPL124}
 O. L. Trinhammer, {\it On Cabibbo angle from theory}, EPL {\bf 124} (2018) 31001.
\bibitem{TrinhammerEPL128}
 O. L. Trinhammer, {\it On energy-momentum tensors and proton structure}, EPL {\bf 128} (2019) 11004.


\end{thebibliography}
\end{document}